\documentclass{article}
\usepackage{spconf,amsmath,graphicx,hyperref}
\usepackage{amssymb}
\usepackage{subcaption}
\usepackage{balance}
\usepackage{tikz}

\allowdisplaybreaks

\def\numAnt{L}
\def\numTrans{G}
\def\trans{g}
\def\numSymbols{K}
\def\symbol{k}
\def\f{\mathbf{f}}
\def\F{\mathbf{F}}
\def\Fopt#1{\mathbf{F}^{\star}\left(#1\right)}

\def\X{\mathbf{X}}
\def\complex{\mathbb{C}}
\def\real{\mathbb{R}}
\def\stV#1#2#3{\mathbf{a}^{#1}\left(#2, #3\right)}
\def\dstV#1#2#3{\mathbf{\dot{a}}^{#1}\left(#2, #3\right)}
\def\r{\mathbf{r}}
\def\maxD{D}

\def\uncerReg{\mathcal{P}}
\def\thSCC{\kappa_{\text{SCC}}}
\def\mainLobeReg{\mathcal{L}}

\title{Optimal Placement of Movable Antennas for Angle-of-Departure Estimation Under User Location Uncertainty}

\name{ \begin{tabular}{c}
     Luc\'ia Pallar\'es-Rodr\'iguez$^{\ast}$\qquad Angelo Coluccia$^{\star}$\qquad Alessio Fascista$^{\S}$\qquad Musa Furkan Keskin$^{\dagger}$ \\  Henk Wymeersch$^{\dagger}$  \qquad Jos\'e A. L\'opez-Salcedo$^{\ast}$\qquad Gonzalo Seco-Granados$^{\ast}$
\end{tabular} \thanks{This work is supported by the Catalan Agency for Management of University and Research Grants (AGAUR) under grant 2024FI-100186, the Spanish Agency of Research (AEI) under grant PID2023-152820OB-I00 funded by MICIU/AEI/10.13039/501100011033, the AGAUR-ICREA Academia Program, and by ERDF/EU and the Swedish Research Council (VR) through the project 6G-PERCEF under Grant 2024-04390.
}}

\address{$^{\ast}$ Department of Telecommunications and Systems Engineering, Universitat Autònoma de Barcelona, Spain \\
      $^{\star}$ Department of Innovation Engineering, Università del Salento, Italy \\
      $^{\S}$ Department of Electrical and Information Engineering, Politecnico di Bari, Italy \\
      $^{\dagger}$ Department of Electrical Engineering, Chalmers University of Technology, Sweden}

\begin{document}
%
\twocolumn[
\begin{@twocolumnfalse}
{\Large{\textbf{IEEE Copyright Notice}}}

\vspace{1em}
\textcopyright 2026 IEEE.  Personal use of this material is permitted.  Permission from IEEE must be obtained for all other uses, in any current or future media, including reprinting/republishing this material for advertising or promotional purposes, creating new collective works, for resale or redistribution to servers or lists, or reuse of any copyrighted component of this work in other works.
\vspace{1em}
\end{@twocolumnfalse}
]
\clearpage
\maketitle
\begin{abstract}
Movable antennas (MA) have gained significant attention in recent years to overcome the limitations of extremely large antenna arrays in terms of cost and power consumption. In this paper, we investigate the use of MA arrays at the base station (BS) for angle-of-departure (AoD) estimation under uncertainty in the user equipment (UE) location. Specifically, we (i) derive the theoretical performance limits through the Cram\'er-Rao bound (CRB) and (ii) optimize the antenna positions to ensure robust performance within the UE's uncertainty region. Numerical results show that dynamically optimizing antenna placement by explicitly considering the uncertainty region yields superior performance compared to fixed arrays, demonstrating the ability of MA systems to adapt and outperform conventional arrays.
\end{abstract}
\begin{keywords}
Movable antenna (MA), antenna position optimization, angle-of-departure estimation, Cram\'er-Rao bound (CRB).
\end{keywords}
\section{Introduction}
\label{sec:intro}

The spatial resolution required for next-generation networks calls for extremely large antenna arrays, increasing hardware cost and power consumption. In this context, movable antenna (MA) configurations have emerged as a promising solution, overcoming these challenges through flexible antenna positioning \cite{zhu24, zhu25}. This technology, also known as fluid antenna systems (FAS) \cite{wong20, new25}, enables real-time adjustment of the antenna positions within a predefined region, improving spatial diversity and angular resolution while reducing spatial correlation among steering vectors from different directions. Unlike fixed antenna architectures offering similar resolutions, such as sparse arrays \cite{wang18}, MAs can be reconfigured dynamically depending on the operating conditions, enhancing robustness to channel variations.

Previous studies have demonstrated the benefits of MAs in the wireless communications domain. In \cite{zhu23}, an uplink multiple-access channel with single-MA user equipment (UE) and a fixed-array base station (BS) is studied, minimizing total transmit power under rate constraints by jointly optimizing MA positions, user power and BS combining matrix. Similarly, \cite{qin24} addresses the downlink channel, optimizing UE MA positions and BS beamforming to minimize transmit power. The authors in \cite{zhu23_2} jointly optimize the positions of an MA array and the weight vector to maximize gain in the desired direction while suppressing interference. In \cite{zhou24}, the use of MAs is extended to the integrated sensing and communications (ISAC) paradigm, and in \cite{an25}, MAs are exploited for beampattern synthesis in spectrum-constrained radar. The potential of MAs for angle-of-arrival (AoA) estimation is explored in \cite{wenyan24}, showing improved angular resolution through optimized MA placement.

In this work, we extend the use of MA systems to angle-of-departure (AoD) estimation, analyzing the impact of antenna placement at the BS when the UE lies within an uncertainty region. We derive the theoretical lower bound based on the Cram\'er-Rao bound (CRB) and incorporate UE location uncertainty into the problem formulation. Due to its complexity, the optimal placement is found through an exhaustive numerical analysis. Results show that optimizing MA positions for the uncertainty region improves AoD estimation, proving the advantages of MA systems over fixed arrays.

\section{System Description}
\label{sec:sys_model}
\subsection{Signal Model}
Let us consider a downlink scenario with a BS equipped with an $\numAnt$-element MA array, and a single antenna UE. The positions of the MAs can be dynamically adjusted along a one-dimensional segment of length $\maxD$, subject to a minimum inter-element spacing $d$. The BS transmits a pilot sequence of $\numSymbols$ symbols repeatedly over $\numTrans$ transmissions, where, at the $\trans$-th transmission, each symbol is precoded by vector $\f_\trans \in \complex^{\numAnt\times 1}$. Assuming far-field and narrowband conditions, the received baseband signal at the UE for the $\symbol$-th symbol during the $\trans$-th transmission is given by
\begin{equation}
    y^k_g = \rho e^{j\varphi}\stV{T}{\theta}{\r}\f_g s^k +\nu^k_g, 
    \label{eq:sig_model}
\end{equation}
where $\rho$ and $\varphi$ denote the magnitude and phase of the complex channel gain, respectively, and $s^k$ is the pilot symbol with average power $P=\frac{1}{\numSymbols}\sum_{k=1}^{\numSymbols} |s^k|^2$. The noise component $\nu^k_g$ follows a Gaussian distribution with zero mean and variance $\sigma^2$. The spatial signature of the BS is characterized by the steering vector $\stV{}{\theta}{\r} \in \complex^{\numAnt\times 1}$, with the $l$-th entry defined as \(\left[\stV{}{\theta}{\r}\right]_l = \exp\left(j\frac{2\pi}{\lambda}\sin(\theta)\left[\r\right]_l\right)\), where $\theta$ denotes the AoD, and $\r\in \real^{\numAnt \times 1}$ is the antenna position vector (APV). The $l$-th element $[\r]_l \in\left[-D/2, D/2\right]$ specifies the position of the $l$-th MA along the array axis, relative to its geometrical center.

\subsection{Theoretical Bound on AoD Estimation}

To analyze the impact of the MAs' position on AoD estimation, we compute the CRB. The unknown channel parameters are the amplitude and phase of the channel gain, as well as the AoD. Under this model, the CRB for the AoD is obtained using the Slepian-Bangs formula \cite{kay} as
\begin{multline}
    \text{CRB}(\r; \theta)=\left[2\,\text{SNR} \left(\vert\vert \F^H\dstV{*}{\theta}{\r}\vert\vert^2  \right.\right.\\\left.\left. -\frac{\vert \stV{T}{\theta}{\r}\F\F^H\dstV{*}{\theta}{\r}\vert^2}{\vert\vert \F^H\stV{*}{\theta}{\r}\vert\vert^2} \right)\right]^{-1},
    \label{eq:CRB_entire}
\end{multline}
where $\text{SNR} = \frac{\numSymbols P\rho^2}{\sigma^2}$ is the signal-to-noise ratio (SNR) at the UE, $\F=\left[\f_1\ \ldots\ \f_\numTrans\right] \in \complex^{\numAnt\times \numTrans}$ is the precoding matrix that ensures $\text{tr}\left(\F\F^H\right) = 1$, and $\dstV{*}{\theta}{\r}\in\complex^{\numAnt\times 1}$ is the derivative of the steering vector with respect to the AoD, with $l$-th entry defined as \(\left[\dstV{}{\theta}{\r}\right]_l = j\frac{2\pi}{\lambda}\cos(\theta)[\r]_l \exp\left(j\frac{2\pi}{\lambda}\sin(\theta)\left[\r\right]_l\right)\).

It is clear from (\ref{eq:CRB_entire}) that the CRB depends not only on $\theta$ and $\r$, but also on the precoding matrix $\F$. Assuming perfect knowledge of the channel parameters, the structure of the optimal precoding matrix can be obtained as the one that minimizes the CRB expression in (\ref{eq:CRB_entire}). To this end, we define $\X = \F \F^H \in \complex^{\numAnt\times\numAnt}$, and we rewrite the CRB as
\begin{multline}
    \text{CRB}(\r, \X; \theta) =\left[2\,\text{SNR}  \left(\dstV{T}{\theta}{\r}\X\dstV{*}{\theta}{\r}  \right.\right.\\\left.\left. -\frac{\vert \stV{T}{\theta}{\r}\X\dstV{*}{\theta}{\r}\vert^2}{\stV{T}{\theta}{\r}\X\stV{*}{\theta}{\r}} \right)\right]^{-1}.
    \label{eq:CRB_propto}
\end{multline}

The dependence of (\ref{eq:CRB_propto}) on $\X$ is through $\dstV{T}{\theta}{\r}\X\dstV{*}{\theta}{\r}$, $\stV{T}{\theta}{\r}\X\dstV{*}{\theta}{\r}$, and $\stV{T}{\theta}{\r}\X\stV{*}{\theta}{\r}$. Thus, the optimal matrix $\X^{\star}$ that minimizes (\ref{eq:CRB_propto}) can be expressed as a combination of the directional and derivative beams \cite{jian08, furkan22}
\begin{equation}
    \X^{\star}=\begin{bmatrix}
      \frac{\stV{*}{\theta}{\r}}{\vert\vert\stV{}{\theta}{\r}\vert\vert} & \frac{\dstV{*}{\theta}{\r}}{\vert\vert\dstV{}{\theta}{\r}\vert\vert }
    \end{bmatrix} \boldsymbol{\Lambda}(\gamma) \begin{bmatrix}
      \frac{\stV{T}{\theta}{\r}}{\vert\vert\stV{}{\theta}{\r}\vert\vert} \\ \frac{\dstV{T}{\theta}{\r}}{\vert\vert\dstV{}{\theta}{\r}\vert\vert } 
    \end{bmatrix},
    \label{eq:optX}
\end{equation}
 where $\boldsymbol{\Lambda}(\gamma)\in\real^{2\times 2}$ is a diagonal matrix defined as 
\begin{equation}
    \boldsymbol{\Lambda}(\gamma) =\begin{bmatrix}
      \gamma & 0 \\ 0 & 1-\gamma 
    \end{bmatrix},
    \label{eq:powAlloc}
\end{equation}
where $\gamma \in (0, 1)$ is the fraction of the power allocated to the directional beam. According to the previous definitions, the optimal precoding matrix for a given $\theta$ and $\r$ is obtained as
\begin{equation}
    \Fopt{\theta, \r,\gamma}=\begin{bmatrix}
      \frac{\stV{*}{\theta}{\r}}{\vert\vert\stV{}{\theta}{\r}\vert\vert} & \frac{\dstV{*}{\theta}{\r}}{\vert\vert\dstV{}{\theta}{\r}\vert\vert }
    \end{bmatrix}\big(\boldsymbol{\Lambda}(\gamma)\big)^{1/2}.
    \label{eq:optF}
\end{equation}
Note that, although the formulation allows for $\numTrans$ precoding vectors, the optimal precoding matrix lies in the two-dimensional subspace spanned by the steering vector at $\theta$ and its derivative. Hence, for $\numTrans>2$, all transmissions can be constructed using the directional and derivative beams in (\ref{eq:optF}), with total power distributed across transmissions such that $\sum_{\trans=1}^{\numTrans}\! \gamma_{\trans}\! =\! 1$, where $\gamma_{\trans}$ is the power allocated to the $\trans$-th transmission. Substituting (\ref{eq:optF}) into (\ref{eq:CRB_entire}), the CRB reduces to:
\vspace{-10pt}
\begin{multline}
    \text{CRB}\left(\r, \Fopt{\theta, \r, \gamma}; \theta\right) = \left[2\,\text{SNR} (1-\gamma)\right. \\ \times \left.\left(\frac{2\pi}{\lambda}\cos(\theta)\right)^2\!\!\r^T\!\r\right]^{-1}.
    \label{eq:CRB_optF}
\end{multline}

\section{Optimal Antenna Placement}

Given (\ref{eq:CRB_optF}), the optimal position of the MAs is obtained by solving the following minimization problem:
\begin{subequations}\label{eq:optCRB_simp}
\begin{align}
\min_{\r}\ \left[2\,\text{SNR} (1-\gamma)\left(\frac{2\pi}{\lambda}\cos(\theta)\right)^2\!\r^T\r\right]^{-1} \label{eq:optCRB_simp_1}\\ \text{s.t.} \; [\r]_{\numAnt}-[\r]_{1} \leq D,  \label{eq:optCRB_simp_2}\\
 \begin{aligned} [ \r ]_{l}-[\r]_{l-1} \geq d,  \\
l=1, \ldots, \numAnt \end{aligned} \label{eq:optCRB_simp_3}
\end{align}
\end{subequations}
where (\ref{eq:optCRB_simp_2}) and (\ref{eq:optCRB_simp_3}) ensure that the maximum dimension of the MA array is not exceeded and that a minimum inter-element spacing is satisfied, respectively. Optimizing (\ref{eq:optCRB_simp}) is equivalent to solving
\begin{subequations}\label{eq:CRB_maxVar}
\begin{align}
\max_{\r}\ \r^T\r \label{eq:CRB_maxVar_1}\\ \text{s.t.} \ \text{(\ref{eq:optCRB_simp_2}),(\ref{eq:optCRB_simp_3})}, \label{eq:CRB_maxVar_2}
\end{align}
\end{subequations}
which corresponds to the problem in \cite{wenyan24}, since $\r^T\r$ can be interpreted as the variance of vector $\r$, whose mean $\frac{1}{\numAnt}\sum_{l=1}^L [\r]_l$ is zero according to the system model definition. The solution to (\ref{eq:CRB_maxVar}), known as the maximum-variance (MaxVar) solution, results in  dividing the available MAs into two equitable groups and positioning each at one end of the segment, while maintaining a minimum inter-element spacing of $d$ to satisfy all constraints.
 
The previous result is obtained under the assumption that the BS uses the precoding matrix $\Fopt{\theta, \r, \gamma}$. However, this requires perfect knowledge of the UE location, since the AoD $\theta$ is used in (\ref{eq:optF}). In this study, we consider the UE position to be unknown within a certain area, so the precoding matrix is designed to steer the directional and derivative beams toward the center of this uncertainty region. An alternative approach is to build multiple beam pairs covering the region, but this increases the number of transmissions relative to the proposed design, since each additional direction adds two vectors to the precoding matrix. Moreover, inter-element spacings larger than half the signal wavelength may generate high-power secondary lobes within the uncertainty region, potentially degrading AoD estimation due to ambiguities. To control the appearance of these side lobes, we introduce an additional constraint on the spatial correlation coefficient (SCC) \cite{heng82, wang14}, defined as
\begin{equation}
    \text{SCC}(\theta_i, \theta_j, \r) = \frac{\stV{H}{\theta_i}{\r}\stV{}{\theta_j}{\r}}{\vert\vert\stV{}{\theta_i}{\r} \vert\vert\ \vert\vert  \stV{}{\theta_j}{\r}\vert\vert},
    \label{eq:defSCC}
\end{equation}
so that the correlation between the steering vector at the center of the uncertainty region and those corresponding to other AoDs within the region remains below a specified threshold. Note that (\ref{eq:defSCC}) can be interpreted as evaluating the response of a directional precoder pointing at angle $\theta_i$ in the AoD $\theta_j$.

Let $\uncerReg$ denote a set of AoDs covering the uncertainty region of the UE position, and let $\theta_c$ be the AoD corresponding to the center of the uncertainty region. The new constraint on the SCC is defined as
\begin{equation}
\max_{\theta \in \uncerReg \setminus\mainLobeReg(\r)}\ \text{SCC}(\theta_c, \theta, \r) \leq \thSCC,
\label{eq:constSCC}
\end{equation}
where $\mainLobeReg(\r)\triangleq\left\{\theta\in\uncerReg\mid \vert\theta-\theta_c\vert < \frac{\theta_{-3\text{dB}}(\r)}{2}\right\}$ denotes the main-lobe region around $\theta_c$, with $\theta_{-3\text{dB}}(\r)$ being the half-power beamwidth, and $\thSCC$ the SCC threshold. The optimal antenna placement in this scenario is obtained by minimizing the worst-case CRB within the uncertainty region using the precoding matrix $\Fopt{\theta_c, \r, \gamma}$, subject to the constraint in (\ref{eq:constSCC}). In this work, we fix the power allocation coefficient to $\gamma=0.5$, and denote the resulting precoding matrix as $\Fopt{\theta_c, \r}$ for simplicity, leading to the following problem: 
\begin{subequations}\label{eq:optCRB_final}
\begin{multline}
\min_{\r}\ \max_{\theta \in \uncerReg}\ \left[2\,\text{SNR} \left(\vert\vert \left(\Fopt{\theta_c, \r}\right)^H\!\dstV{*}{\theta}{\r}\vert\vert^2  \right.\right.\\\left.\left. -\frac{\vert \stV{T}{\theta}{\r}\Fopt{\theta_c, \r}\!\left(\Fopt{\theta_c, \r}\right)^H\!\dstV{*}{\theta}{\r}\!\vert^2}{\vert\vert \left(\Fopt{\theta_c, \r}\right)^H\!\stV{*}{\theta}{\r}\vert\vert^2} \right)\!\right]^{-1}\! \label{eq:optCRB_final_1}\end{multline}
\begin{align}\text{s.t.} \ \text{(\ref{eq:CRB_maxVar_2}),(\ref{eq:constSCC}}).
\end{align}
\end{subequations}

Solving (\ref{eq:optCRB_final}) is challenging due to its non-convex objective function and constraints. Therefore, to gain insight into its behavior, we perform a simulation-based analysis to characterize the optimal antenna placement and evaluate how it varies for different sizes of the uncertainty region.

\section{Numerical Results}

\subsection{Simulation Setup}

\begin{figure}[htb]
\centering
\resizebox{0.95\linewidth}{!}{\begin{tikzpicture}
\draw[black, very thick,->] (0,0) -- (10.2,0);
\draw[black,  thick,<->] (0,-1) -- (9,-1);
\node at (4.5,-1.3) {$D$};
\draw[black, thick] (4.5,-0.1) -- (4.5,0.1);
\filldraw[black, very thick] (0,0) circle (0.2);
\filldraw[black, very thick] (1.5,0) circle (0.2);
\filldraw[black, very thick] (3,0) circle (0.2);
\filldraw[black, very thick] (6,0) circle (0.2);
\filldraw[black, very thick] (7.5,0) circle (0.2);
\filldraw[black, very thick] (9,0) circle (0.2);
\draw  node at (0,-0.5) {$-r_1$} node at (1.5,-0.5) {$-r_2$} node at (3,-0.5) {$-r_3$} node at (6,-0.5) {$r_3$} node at (7.5,-0.5) {$r_2$} node at (9,-0.5) {$r_1$} node at (4.5,-0.3) {$0$} node at (10, -0.25) {$r$};
\draw[thick, <->] (0,0.5) -- (1.5,0.5);
\node at (0.75, 0.7) {$a$};
\draw[thick, <->] (1.5,0.5) -- (3,0.5);
\node at (2.25, 0.7) {$b$};
\draw[thick, <->] (6,0.5) -- (7.5,0.5);
\node at (6.75, 0.7) {$b$};
\draw[thick, <->] (7.5,0.5) -- (9,0.5);
\node at (8.25, 0.7) {$a$};
\end{tikzpicture}
}
\caption{Configuration of the MA array.}
\label{fig:antConf}
\end{figure}
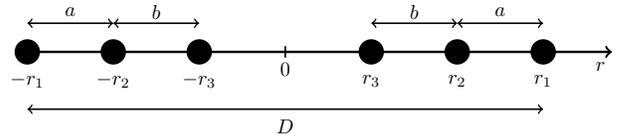

To conduct the numerical study we assume the array is composed of six MAs ($\numAnt=6$), with a maximum dimension of $\maxD = 10\lambda$ and $d=\lambda/2$. The array is parametrized by two variables, $a$ and $b$ ($a, b\in \left[d, (\maxD - 3d)/2\right]$), which control the distance between the antennas, as illustrated in Fig. \ref{fig:antConf}. 

According to Fig. \ref{fig:antConf}, the MaxVar solution places the antennas at $r_1 = \maxD/2$, $r_2 = \maxD/2- d$, and $r_3 = \maxD/2 - 2d$. To determine the solution to (\ref{eq:optCRB_final}), we evaluate the worst-case CRB within the uncertainty region over an exhaustive range of values for $a$ and $b$, and we select the optimal placement as the one that minimizes the expression while satisfying all the constraints. Note that, in all the APVs evaluated, the extremes of the array are fixed to occupy the entire available space, i.e., $r_1=D/2$. To assess the effect of the uncertainty region size on the optimal APV (Opt-APV), we define two regions, $\uncerReg^1 \subseteq [0^\circ, 20^\circ]$ and $\uncerReg^2\subseteq [-10^\circ, 30^\circ]$, with a central angle $\theta_c=10^\circ$ to steer $\Fopt{\theta_c, \r}$. Finally, the $\thSCC$ threshold is set to 0.5.

\subsection{Results and Discussion}

\begin{figure}[tb]
\begin{subfigure}[b]{1.0\linewidth}
  \centering
  \centerline{\includegraphics[clip, trim={0cm 0cm 0cm 0.5cm}, width=8.5cm]{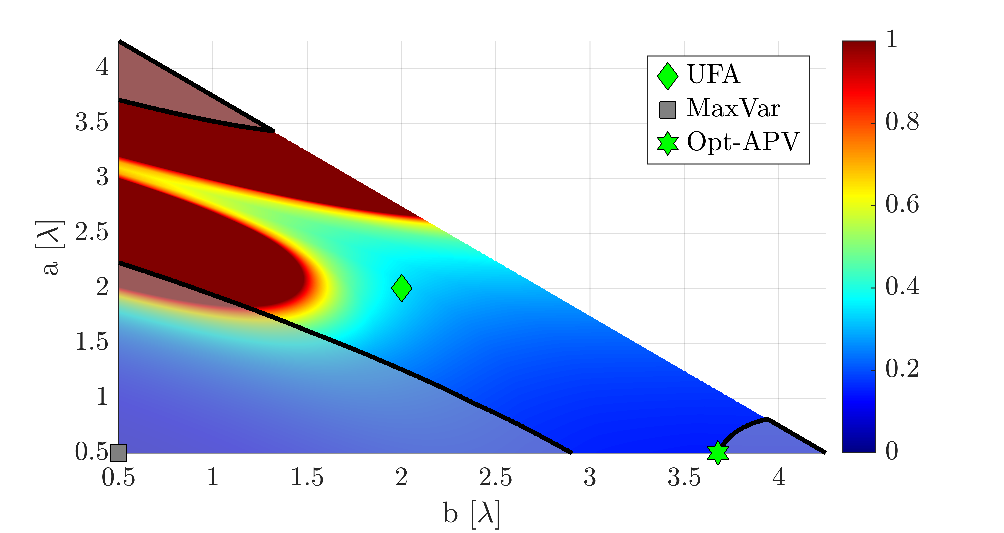}}
  \subcaption{Uncertainty region $\uncerReg^1$.}\medskip
  \label{fig:wCRB_1}
\end{subfigure}

\begin{subfigure}[tb]{1.0\linewidth}
  \centering
  \centerline{\includegraphics[clip, trim={0cm 0cm 0cm 0.5cm}, width=8.5cm]{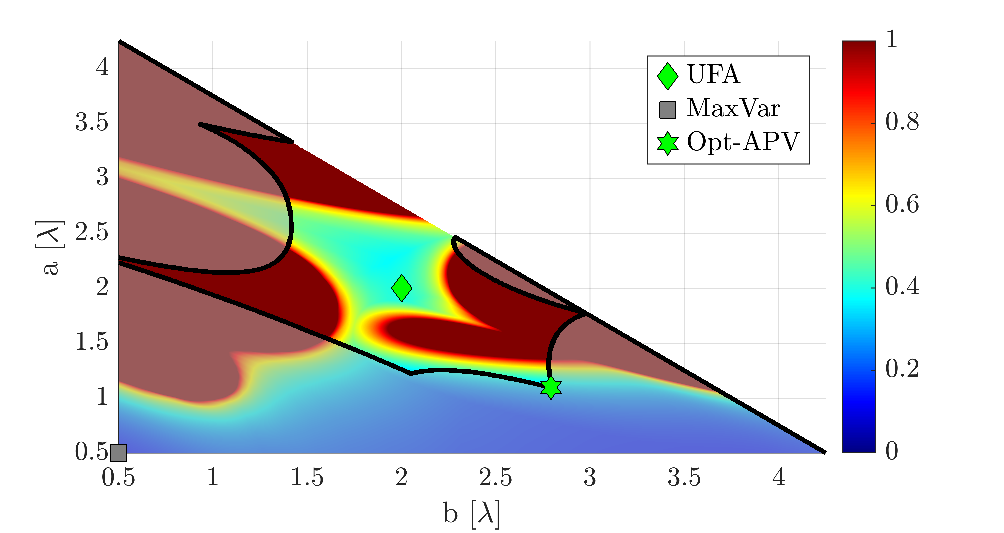}}
  \subcaption{Uncertainty region $\uncerReg^2$.}\medskip
  \label{fig:wCRB_2}
\end{subfigure}
\caption{Worst-case CRB in degrees for (a) $\uncerReg^1$ and (b) $\uncerReg^2$.}
\label{fig:res_1}
\end{figure}

\begin{figure}[tb]
\centering
  \centerline{\includegraphics[width=8.5cm]{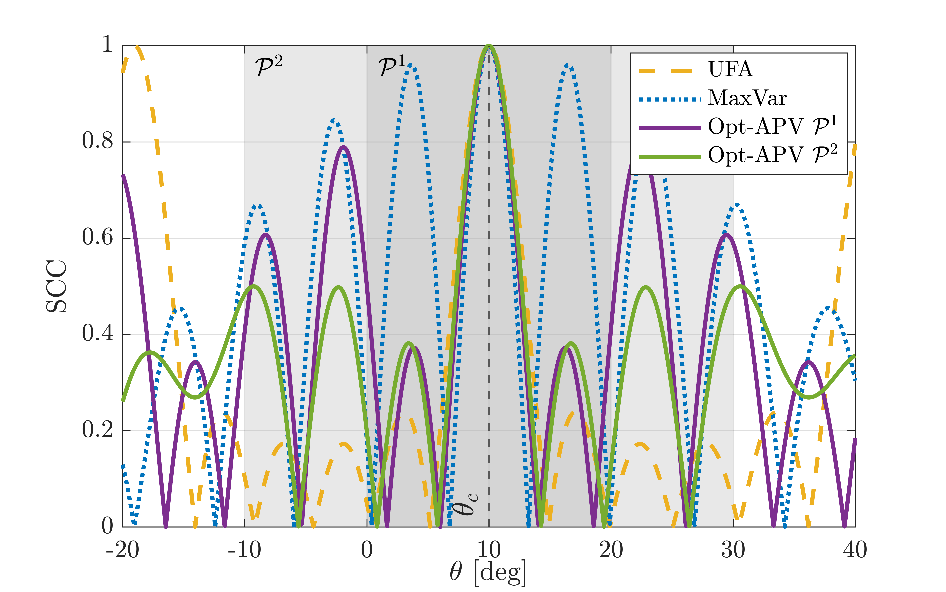}}
 \caption{$\text{SCC}(\theta_c, \theta, \r)$ as a function of $\theta$.}
\label{fig:SCC}
\end{figure}

In Fig. \ref{fig:res_1}, the worst-case CRB is shown for both uncertainty regions over all possible combinations of $a$ and $b$, with the regions where the SCC constraint is not satisfied highlighted in gray. Particularly, Fig. \ref{fig:wCRB_1} shows the results for $\uncerReg^1$, while Fig. \ref{fig:wCRB_2} shows those for $\uncerReg^2$. In both cases, if the SCC constraint were disregarded, the optimal APV would correspond the MaxVar solution. However, Fig. \ref{fig:SCC} shows that the MaxVar APV produces high SCC values within the uncertainty regions, requiring the rearrangement of the MAs to avoid ambiguities. Moreover, for the uniform full-aperture array (UFA), where the antennas are equidistantly spaced along the full dimension, the SCC remains below the threshold, but the worst-case CRB is higher than that achievable with the optimal APV.

As depicted in Fig. \ref{fig:res_1}, increasing the size of the uncertainty region reduces the set of feasible solutions due to the SCC constraint, leading to different optimal APVs in each case. This is clearly illustrated in Fig. \ref{fig:SCC}, where the optimal APV for $\uncerReg^1$ produces SCC values around 0.8 within $\uncerReg^2$, rendering it infeasible for the latter region. Such behavior is of high importance, as it highlights the need to reconfigure the MAs according to the uncertainty region for the UE location, which may change during system operation.

\begin{figure}[tb]
\centering
  \centerline{\includegraphics[width=8.5cm]{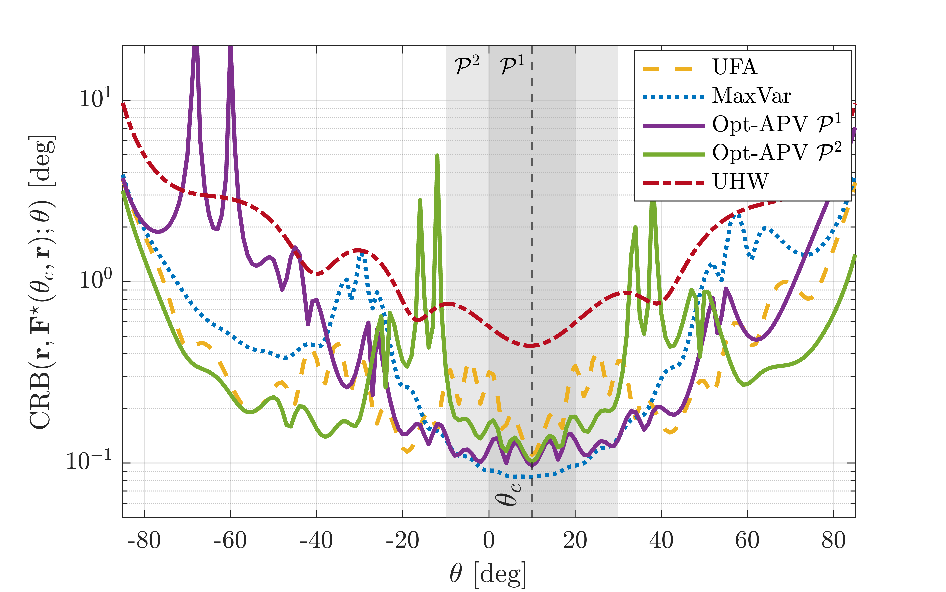}}
 \caption{$\text{CRB}(\r ,\Fopt{\theta_c,\r};\theta)$ as a function of $\theta$.}
\label{fig:CRB_theta}
\end{figure}

In Fig. \ref{fig:CRB_theta}, the CRB for each APV using the optimal precoders is depicted as a function of $\theta$. For comparison, in addition to the optimal APV for each region and the MaxVar and UFA solutions, the standard uniform linear array with half-wavelength inter-element spacing (UHW) is also included. Note that, although the MaxVar solution offers the lowest CRB, the optimal solutions for each region clearly outperform both fixed antenna arrays, the UFA and UHW, while satisfying the SCC constraint. 

\begin{figure}[tb]
\centering
    \centerline{\includegraphics[width=8.5cm]{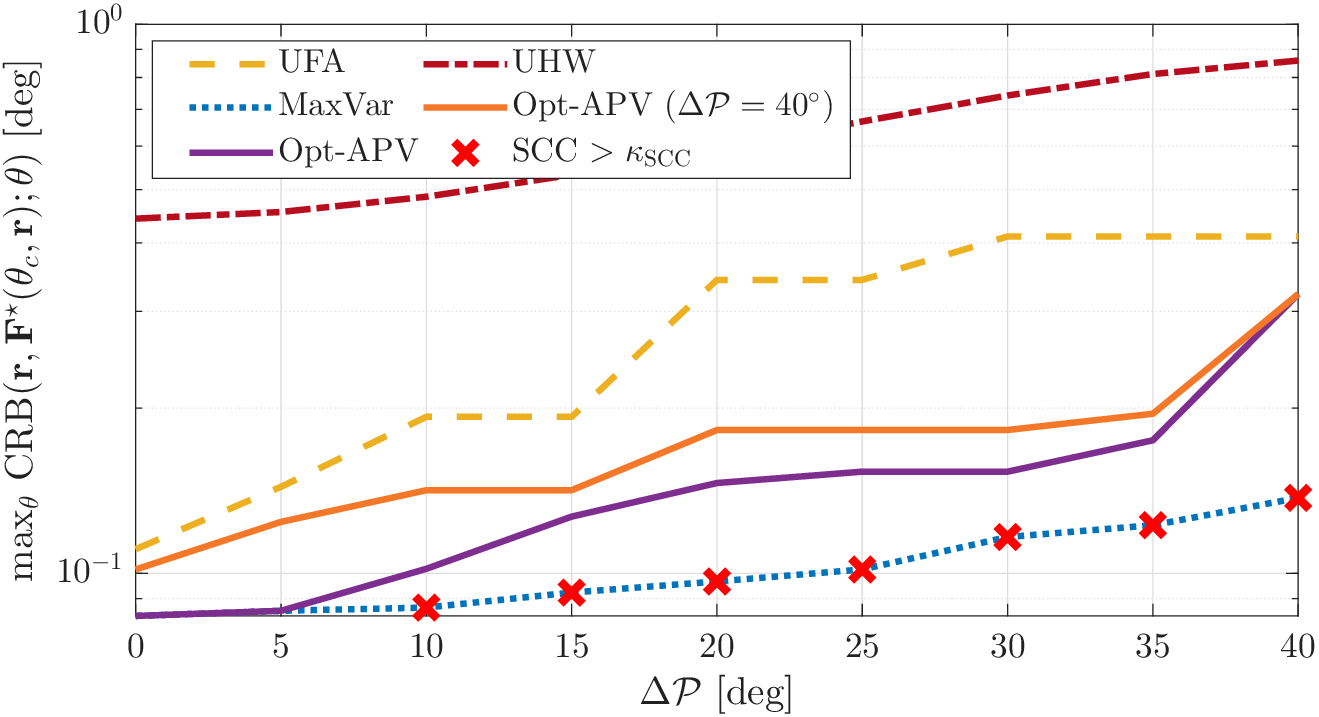}}
 \caption{Worst-case CRB versus uncertainty region size.}
\label{fig:wCRB_uncerReg}
\end{figure}

Lastly, to demonstrate the importance of selecting the optimal placement of the MAs according to the uncertainty region, Fig. \ref{fig:wCRB_uncerReg} shows the worst-case CRB achieved by each solution for different uncertainty region sizes $\Delta\uncerReg$, where $\Delta\uncerReg = \max\left(\uncerReg\right)-\min\left(\uncerReg\right)$. For regions covering less than $10^\circ$, the Opt-APV corresponds to the MaxVar solution, since no grating lobes appear within such small regions. As expected, as the angular span increases, the Opt-APV solution provides the lowest worst-case CRB among the solutions satisfying the SCC constraint, surpassing both the UFA and UHW for all $\Delta\uncerReg$. In addition to the previously considered APVs, we evaluate the behavior of the Opt-APV corresponding to the largest region (Opt-APV $(\Delta\uncerReg=40^\circ)$) when applied to all region sizes. These results further highlight the advantages of MAs, as the APV can be continuously optimized according to the system's operating conditions, adjusting the antenna positions to account for the varying uncertainty in the UE location.

\section{Conclusion}

In this work, we investigated the optimal placement of MAs for AoD estimation under uncertainty in the UE location. We developed theoretical bounds on AoD estimation and analyzed the impact of introducing an uncertainty region into the problem. To address the resulting non-convex optimization problem, we conducted a numerical study to determine the optimal APV. The results show that adapting the APV according to the uncertainty region significantly reduces the worst-case CRB compared to fixed antenna configurations, highlighting the role of optimal antenna placement in achieving low AoD estimation errors.



\balance 
\bibliographystyle{IEEEbib}
\bibliography{refs}
\label{sec:refs}
\end{document}